\documentclass{article}
\usepackage[utf8]{inputenc}
\usepackage{authblk}
\usepackage{svg}
\usepackage{graphicx}
\usepackage{biblatex}
\usepackage[font=small,margin=20pt]{caption}
\usepackage{subcaption}

\usepackage{cleveref}
\Crefname{figure}{Fig.}{Figs.}% {<type>}{<singular>}{<plural>}

\usepackage{xcolor}

\title{Infomap Bioregions 2 - Exploring the interplay between biogeography and evolution}

\author[*,1,2,3]{Daniel Edler}
\author[1]{Anton Holmgren}
\author[1,4]{Alexis Rojas}
\author[5]{Joaquín Calatayud}
\author[1]{Martin Rosvall}
\author[2,3,6,7]{Alexandre Antonelli}

\affil[1]{{\small Integrated Science Lab, Department of Physics, Umeå University, Ume\aa, Sweden}}
\affil[2]{{\small Gothenburg Global Biodiversity Centre, University of Gothenburg, Gothenburg, Sweden}}
\affil[3]{{\small Department of Biological and Environmental Sciences, University of Gothenburg, Gothenburg, Sweden}}
\affil[4]{{\small Department of Computer Science, University of Helsinki, Finland}}
\affil[5]{{\small Department of Biology, Geology, Physics and Inorganic Chemistry, King Juan Carlos University, Madrid, Spain}}
\affil[6]{{\small Department of Biology, University of Oxford, Oxford, United Kingdom}}
\affil[7]{{\small Royal Botanical Gardens Kew, Richmond, Surrey, United Kingdom}}
\affil[*]{{\small Corresponding author: daniel.edler@umu.se}}

\begin{document}

\maketitle

\begin{abstract}

Identifying and understanding the large-scale biodiversity patterns in time and space is vital for conservation and addressing fundamental ecological and evolutionary questions. Network-based methods have proven useful for simplifying and highlighting important structures in species distribution data. However, current network-based biogeography approaches cannot exploit the evolutionary information available in phylogenetic data.
We introduce a method for incorporating evolutionary relationships into species occurrence networks to produce more biologically informative and robust bioregions. To keep the bipartite network structure where bioregions are grid cells indirectly connected through shared species, we incorporate the phylogenetic tree by connecting ancestral nodes to the grid cells where their descendant species occur. To incorporate the whole tree without destroying the spatial signal of narrowly distributed species or ancestral nodes, we weigh tree nodes by the geographic information they provide. For a more detailed analysis, we enable integration of the evolutionary relationships at a specific time in the tree. By sweeping through the phylogenetic tree in time, our method interpolates between finding bioregions based only on distributional data and finding spatially segregated clades, uncovering evolutionarily distinct bioregions at different time slices. We also introduce a way to segregate the connections between evolutionary branches at a selected time to enable exploration of overlapping evolutionarily distinct regions. We have implemented these methods in Infomap Bioregions, an interactive web application that makes it easy to explore the possibly hierarchical and fuzzy patterns of biodiversity on different scales in time and space.

\end{abstract}

\clearpage

\section{Introduction}

Biogeographical regions, bioregions for short, disclose the organization of life. They reveal how species are spatially grouped at large spatial scales and are important units for understanding historical biogeography, ecology, and evolution. They also identify units of biodiversity with potentially high conservation value~\cite{vynne_ecoregion-based_nodate}.

There are two main data-driven approaches to identifying bioregions from spatial species occurrence data: similarity-based\cite{laffan_biodiverse_2010,kreft_framework_2010} and network-based~\cite{vilhena_network_2015,edler_infomap_2017}.
% PARAGRAPH THAT EXPLAINS HOW SIMILARITY-BASED APPROACHES WORK AND WHY THEY FAIL.
%
Similarity-based approaches group grid cells with similar sets of species. Some generalizations include phylogenetic similarity between species~\cite{duarte_dissecting_2016,maestri_evoregions_2020,liu_updated_2023}. However, while similarity-based methods are simple to interpret, they require all grid cells to be similar in a bioregion. Therefore, they cannot capture regions with a spatial gradient of species -- where each pair of neighboring grid cells have a large overlap of species but two grid cells further apart have low overlap -- often with arbitrary boundaries somewhere along the gradient~\cite{vilhena_network_2015}. They also require choosing key parameters that can strongly influence the results, such as similarity measures, thresholds, and tie-breaking~\cite{bloomfield_comparison_2018}. Therefore, researchers have looked for other approaches to find robust and biologically realistic bioregions.

Network-based methods uncover bioregions from species distribution data by first binning all species into spatial grid cells, typically one-by-one degree, which is approximately 100-by-100~km at the Equator, or finer scales when there is sufficient data.
Then, grid cells and species are represented as a bipartite network, with links between species and grid cell nodes specifying occurrences. A link signifies that a species has been recorded in the cell, typically by a collected and geo-referenced specimen in a natural history collection or based on verified observations. In most cases, such a network will have a modular structure: nodes with a relatively high density of links among themselves form communities, identified with community-detection algorithms.
In species occurrence networks, communities correspond to areas with more shared species within than between them. Each community contains grid cells and species, and the grid cells assigned to the same community form a bioregion. One can project the bipartite network to a unipartite network consisting of only grid cells by replacing links to the same species with a direct link between grid cells. Such a projection can also provide interpretable communities but seem to distort information in the underlying data~\cite{rojas_multiscale_2021}.

Network-based methods overcome the similarity-based gradient problem\cite{vilhena_network_2015}. They recognize that occupancy histories of species exhibit varying shapes\cite{foote_rise_2007} and thus, individual species have a different contribution to the overall biogeographic structure. However, current network approaches still ignore the evolutionary relations between species by treating two species within the same genus and two from different families, regardless of their unique evolutionary histories. This has potential consequences for the network-delineated biogeographic structure. For example, common differences in taxonomic classification based on different views or methods to determine which groups of organisms should be considered the same or different species (the so-called lumper-splitter problem) can harm the result~\cite{faurby_strong_2016}, and sparse occurrence data may lead to fragmented maps. %Species numbers for even charismatic taxa, such as primates, vary largely in the literature (from c. x to y species), despite the fact that the group's phylogenetic tree is uncontested.
By discarding vital biodiversity knowledge on evolutionary relationships and relying on fluid species boundaries, we limit bioregions' value for understanding large-scale biodiversity patterns and conservation planning.

To overcome these limitations, we introduce a method for incorporating evolutionary relationships into species occurrence networks to produce more biologically realistic, data-rich, and robust bioregions. %Our method better uses available biodiversity knowledge and helps to identify more effective conservation targets.
%PARAGRAPH THAT SUMMARIZES THE RESULTS.
We find that integrating information about the phylogenetic tree uncovers evolutionarily more distinct and robust bioregions.

\section{Material and Methods}
%To find biologically realistic regions, we include evolutionary relations between species.
To delineate bioregions from species occurrence data, we represent the data as a network with links between species and the grid cells where they occur. Using a network community detection algorithm, we partition the network into optimal modules (Fig.~\ref{fig:demo1a}). Bioregions are defined as grid cells belonging to the same module. As the network contains two types of nodes and links always connect species to grid cells, never species to species or grid cells to grid cells, the network is bipartite. The bipartite structure is essential for modules to be interpreted as areas with more shared species within than between them. Here we extend this method to detect bioregions not only based on shared species but also shared ancestry.
%Either at a selected time or from the full evolutionary history weighted by the spatial extent of the descendants to each internal node in the phylogenetic tree.

The evolutionary relations between species are represented with a fully dichotomous, ultrametric tree structure derived from a phylogenetic inference. Branch lengths are normally provided as absolute time resulting from a divergence time analysis but could be relative to the total tree height.

We introduce a way to integrate the full phylogenetic tree and two complementing ways of including the tree at a specific time for further analysis.

\subsection*{Integrating the whole phylogenetic tree}

To form bioregions connected not only by shared species but by shared evolutionary histories, we connect each ancestral node in the phylogenetic tree to the grid cells where its descendant species occur. As ancestral nodes further back in time connect to more and more grid cells, they will quickly make the network too dense to find any modules. To solve this, we down-weight links from ancestral nodes based on how much spatial information they give (Fig.~\ref{fig:demo1b}). If tree node $i$ is connected to $k_i$ grid cells out of $K$ total grid cells, we weigh the links from tree node $i$ to all connected grid cells by $$
w_i = 1 - \frac{\log k_i}{\log K}.
$$
This gives zero weight to ancestral nodes that exist in all grid cells and weight $1$ to all that exists in a single grid cell. As we do not weigh species by abundances in each grid cell, the $\log k_i$ can be interpreted as the entropy or uncertainty in knowing from which grid cell a selected species occurrence comes, with $\log K$ being the max uncertainty used for normalization.

%In this manuscript we use a dataset with globally distributed point occurrences of mammals

%The evolutionary tree can be thought of as a network of its own, but lacking the bipartite structure found in the species-grid cells network. If one were to connect the evolutionary tree network with the related species in the species-grid cell network, the resulting structure will no longer be bipartite as connections can be species-grid cell, species-genus, genus-family, and so on. This means that a community detection algorithm can find modules that do not contain grid cells. We need to integrate the evolutionary tree with the species-grid cell network without losing the bipartite structure to ensure that the found modular structure represents bioregions.
%If we add the tree to the bipartite network, the ancestral nodes will not directly be connected to any grid cells, and we may end up with modules that capture parts of the tree structure without any grid cells.

\subsection*{Integrating ancestral nodes for mapping the evolution of bioregions}

\begin{figure}[htbp]
     \begin{subfigure}[t]{0.48\textwidth}
         \centering
         \includegraphics[width=\textwidth]{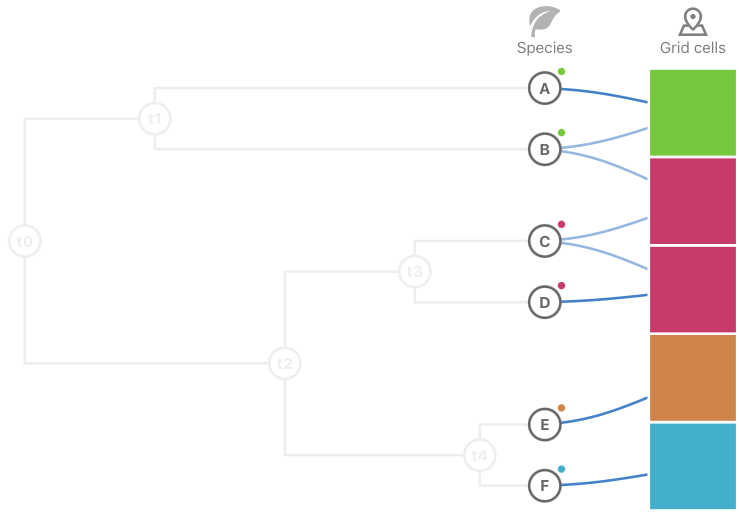}
         \caption{}
         \label{fig:demo1a}
     \end{subfigure}
     \hfill
     \begin{subfigure}[t]{0.48\textwidth}
         \centering
         \includegraphics[width=\textwidth]{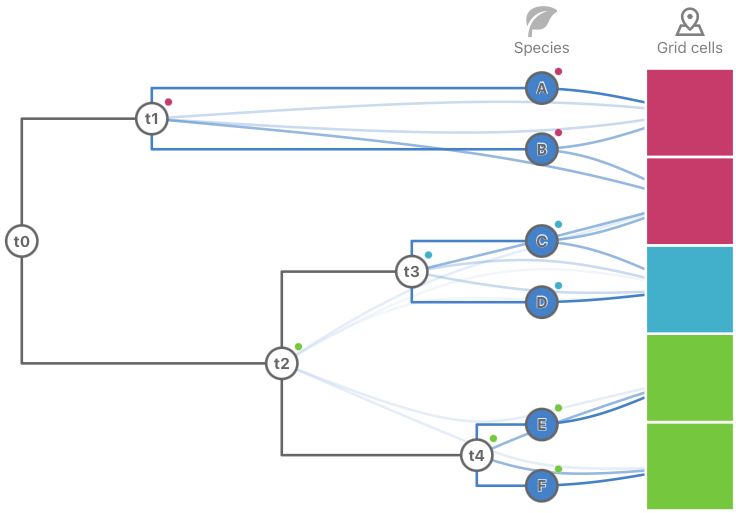}
         \caption{}
         \label{fig:demo1b}
     \end{subfigure}
     \begin{subfigure}[t]{0.48\textwidth}
         \centering
         \includegraphics[width=\textwidth]{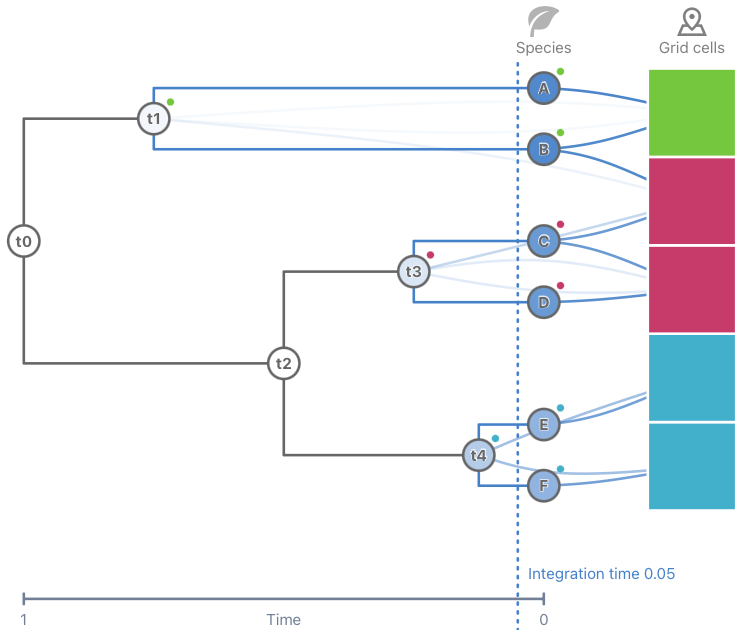}
         \caption{}
         \label{fig:demo1c}
     \end{subfigure}
     \hfill
     \begin{subfigure}[t]{0.48\textwidth}
         \centering
         \includegraphics[width=\textwidth]{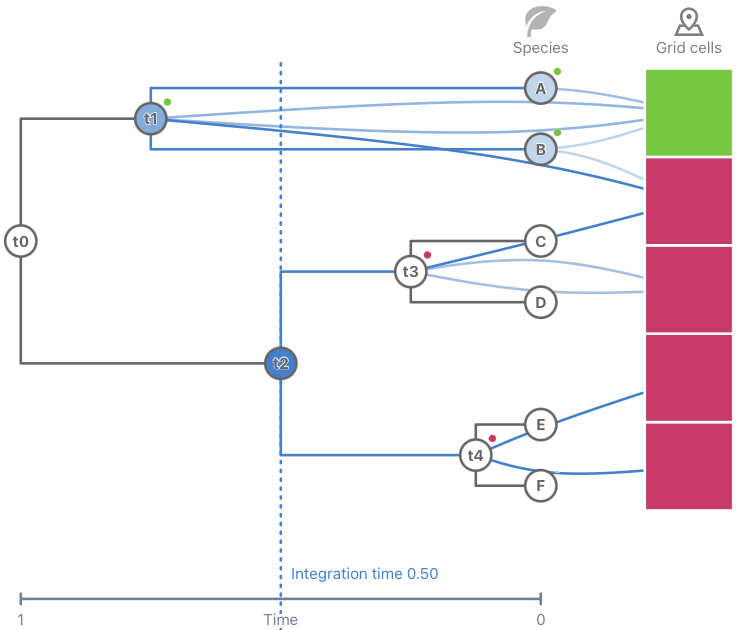}
         \caption{}
         \label{fig:demo1d}
     \end{subfigure}
    \caption{Community detection in bipartite occurrence networks including evolutionary relationships. To find biogeographical regions that capture the spatial structuring of species, we apply community detection on the bipartite network composed of species linked to grid cells where they occur. Grid cells are indirectly connected through shared species (a). The two bottom grid cells end up in different bioregions because of no shared species. By connecting each ancestral node in a phylogenetic tree to the grid cells where its descendant species occur, weighted by their spatial information, we can form bioregions connected not only by shared species but by shared evolutionary history (b). We can indirectly connect grid cells by their shared ancestry at a selected time (c)-(d) for a more detailed analysis. With a recent time, we can use it to solve fragmented bioregions due to sparse data and make it robust against shifting taxonomic resolutions due to the so-called lumper-splitter problem. We uncover unique bioregions by sweeping the selected integration point back in time.}
    \label{fig:demo1}
\end{figure}

To explore the effect of integrating species relationships in more detail to uncover evolutionarily unique bioregions, we introduce a way to integrate the tree at a selected point in time (Fig.~\ref{fig:demo1}c-d). 
To keep a constant weight from the tree as we explore different selected times of integration, we do not weigh links from ancestral nodes by their geographic information but instead by a selected relative tree strength compared to the species nodes of the network without any tree data. The aim is to sweep through the selected time of integration from recent to old while keeping a selected fixed balance between the strength of the ancestral nodes and the strength of the species nodes. As older ancestral nodes are more well-connected, the links from ancestral nodes will be down-weighted to keep a fixed total weight.    
However, any specific time point usually cuts the tree's branches instead of landing at specific ancestral nodes. Below we describe how to solve this by splitting the link weights between the parent and child nodes of each cut branch such that the relative strength between them matches their relative distance to the selected time, with the strength of a node being the sum of link weights connected to that node.

As a parent node may be connected to more grid cells than the child node, we cannot directly split the weights using the relative time between them.
When a branch from parent to child is cut at a relative time $t$ going from $0$ at the parent to $1$ at the child, we weigh the links linearly with $t$.
If the number of links of the parent and child node is $k_p$ and $k_c$, the strength of the parent and child nodes are $s_p = k_p (1-t)$ and $s_c = k_c t$ respectively.
With the relative degree $k_{p/c} = k_p/k_c$, the relative strength of the parent node at $t$ is 
$$ s_{p/c}(t) = \frac{k_p (1-t)}{k_c t} = k_{p/c}\frac{(1-t)}{t}. $$
To interpolate linearly between parent and child strength, we can use an adjusted time $t'$ to cancel the bias $k_{p/c}$
$$ t' = \frac{t k_{p/c}}{1 + t (k_{p/c} - 1)} $$
such that
$$ k_{p/c}\frac{(1-t')}{t'} = \frac{(1-t)}{t}. $$
After weighting the links for parent and child nodes with $1 - t'$ and $t'$, the total strength of all included tree nodes is $s_t$.
With the total strength from the species nodes without the tree $s_s$, we rescale all links from the tree nodes with a factor $w_t$.
This gives the total strength $s'_t = w_t s_t$ such that the relative strength of the tree is the specified value $s_{t/s} = s'_t / (s'_t + s_s)$.
Solving for $w_t$ gives
$$ w_t = \frac{s_{t/s} s_s}{s_t (1 - s_{t/s})}. $$

While this approach uncovers evolutionarily largely unique bioregions, such as islands or mountain ranges,
the typical spatial overlap between evolutionary branches amplifies the problem in current network-based methods to handle evolutionary transition zones with overlapping ecosystems.
As grid cells are forced to be part of only a single bioregion, grid cells in a transition zone are either grouped to their bioregion, arbitrarily assigned to one of the neighboring regions, or collapsed with the neighboring regions into one bioregion, leading to underfitting.
As we extend the integration time, we include ancestral nodes with increasing spatial overlap. This tends to collapse modular structures.
To overcome this, we need a way to find overlapping bioregions. % which we achieve through a segregation approach as described below.

\subsection*{Segregating evolutionary history for mapping overlapping evolutionarily distinct bioregions}

%\begin{figure}[htbp]
%  \centering
%  \includegraphics[width=\columnwidth]{figures/demo.png}
%  \caption{Community detection in bipartite occurrence networks including evolutionary relationships. To find biogeographical regions that capture the spatial structuring of species, we apply community detection on the bipartite network composed of species linked to grid cells where they occur. To find evolutionarily distinct regions, we include phylogenetic relationships between species. We include those by adding ancestral nodes at a selected integration time to the network and linking them to the grid cells where their descendant species occur. a) The bipartite network of species and grid cells without incorporating the tree. The two bottom grid cells end up in different bioregions because of no shared species. b) By moving the integration time back, we connect the two bottom grid cells through a recent common ancestor of species E and F, reducing fragmented bioregions for sparse data. c) Moving the integration time further back merges more grid cells and highlights the spatial distribution of the two major branches of the tree. However, they overlap on the second to top grid cell, which may end up in any of the two bioregions depending on the random seed. d) A higher-order network, using state nodes within grid cells to remember from which ancestral branch at selected segregation time as memory nodes within the grid cells. This allows overlapping bioregions.}
%  \label{fig:schematic}
%\end{figure}

%Speciation has a strong influence on the geographic patterns of species.
By integrating tree nodes further and further back in time, we retain evolutionarily unique bioregions and map selected divergence events. However, evolutionary branches often overlap spatially, so ancestral nodes increase the indirect links between bioregions, eventually collapsing them into large regions (Fig.~\ref{fig:demo1d}).

Using the map equation framework for network community detection~\cite{edler_mapequation_2022}, we can find overlapping evolutionary distinct bioregions using higher-order networks~\cite{edler_mapping_2017}.
The map equation is based on a dynamic approach to community detection, where modules capture the flow patterns on top of a network -- modeled by the relative abundance of stationary visit rates of a random walk.
The random walker moves between nodes, at each step moving to a random neighboring node proportional to the weight of that link.
As the probability distribution for the next node only depends on the current node, this is a memoryless first-order Markov process.
Higher-order Markov processes depend on one or more previous nodes in the sequence.
A higher-order model in the map equation framework is represented by a first-order random walk on state nodes, where each state node belongs to one physical node.
A state node represents the memory of previous steps.
A physical node can overlap modules by partitioning the state nodes into different modules.

\begin{figure}[htbp]
     \centering
     \begin{subfigure}[t]{0.48\textwidth}
         \centering
         \includegraphics[width=\textwidth]{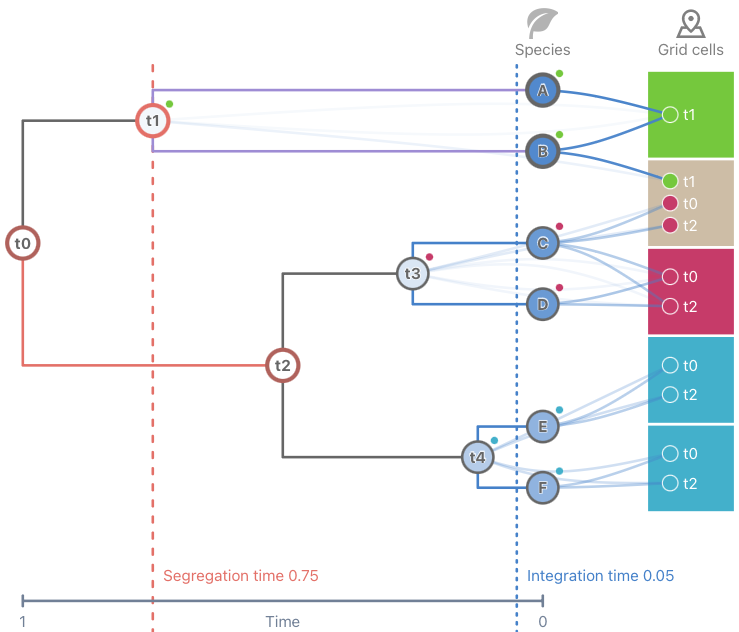}
         \caption{}
         \label{fig:demo2a}
     \end{subfigure}
     \hfill
     \begin{subfigure}[t]{0.48\textwidth}
         \centering
         \includegraphics[width=\textwidth]{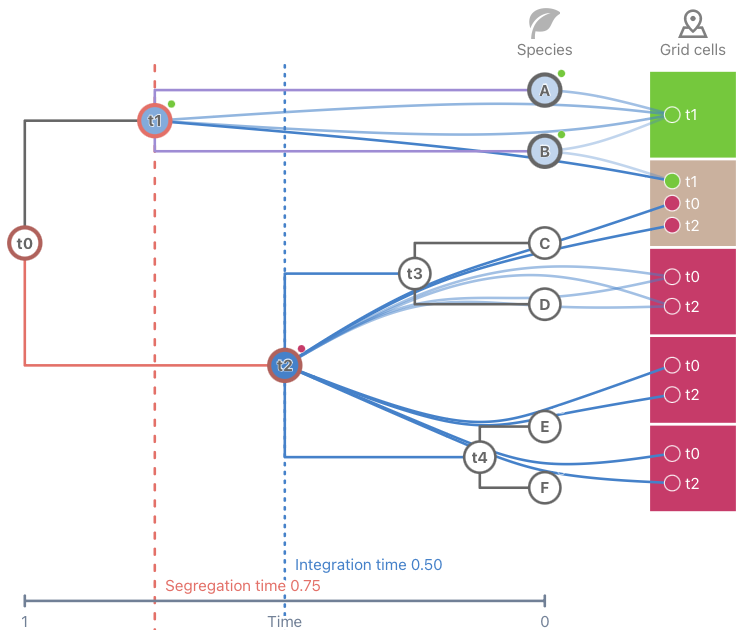}
         \caption{}
         \label{fig:demo2b}
     \end{subfigure}
    \caption{Mapping overlapping evolutionary distinct bioregions. We use a higher-order network with state nodes within grid cells to segregate connections to species within the same evolutionary branch up to a selected time. State nodes represent evolutionary memory: when we follow a link from a species or ancestral node to a grid cell, it forces us to go back to a random species or ancestral node with shared memory. On a state-node level, this gives evolutionarily distinct bioregions. On a physical grid-cell level, this enables the bioregions to overlap in space. By combining it with integrating the tree at a selected time, we can explore evolutionarily unique regions with higher resolution, as spatially overlapping branches may be kept segregated.  }
    \label{fig:demo2}
\end{figure}

To distinguish the geographic patterns from distinct branches when connecting species to grid cells, we employ a second-order model on the grid cells to remember which evolutionary branch at a selected time each species belongs.
We represent this evolutionary memory with state nodes within grid cells, one state node for the parent and one for the child node of the branch cut by the selected segregating time (Fig.~\ref{fig:demo2}).
Each link is then divided into two links connecting grid cells to species and ancestral nodes, with link weights distributed relative to the time at the cut between parent and child nodes.
When ancestral nodes are included, the links are aggregated from their descendant species on the memory nodes.
If the ancestral nodes are closer in time to the species nodes than the segregating time, all network flows are completely segregated within the respective evolutionary branch at the cut.
By increasing the integration time so that ancestral nodes further away than the segregation time are included, they bridge segregated branches that descend from them and possibly connect nodes from them into the same module.
By detecting communities in the state network, each grid cell's state nodes will belong to a module corresponding to an evolutionarily distinct bioregion.
Each grid cell may have multiple state nodes, so each can belong to multiple overlapping bioregions.

\subsection*{Nested bioregions and transition zones}
A biogeographical transition zone is defined as a geographical area of overlap, with a gradient of replacement and partial segregation between biotic components (sets of taxa that share a similar geographical distribution as a product of a common history)\cite{ferro_biogeographical_2014}.
Many factors, such as mountain ranges, watersheds, climate, macroevolutionary and paleogeographic processes, flora, and fauna, shape bioregional boundaries~\cite{calatayud_pleistocene_2019,ficetola_global_2017,liu_updated_2023}. Some boundaries may be sharp, while others may be fuzzy~\cite{maestri_evoregions_2020,liu_updated_2023}. To highlight transition zones, we use a network metric called the participation coefficient that defines the degree to which a node is connected to other modules~\cite{bloomfield_comparison_2018}. We adopt the measure by coloring grid cells with an opacity equal to the fraction of connected species belonging to the same module, weighted by the link weights to the species. To further enhance the richness of the bioregional maps, we mix the color of each grid cell with the color of the second most connected bioregion, interpolated according to their relative weight.

However, wide-spread species can obscure and collapse modular patterns and transition zones defined by range-restricted species~\cite{laffan_range-weighted_2016}. As narrowly distributed species are important for unveiling biogeographic patterns and evolutionary processes~\cite{quintero_global_2018}, and usually also considered of higher conservation value~\cite{farooq_wege_2020}, we down-weight widespread species as described above for integrating the whole phylogenetic tree.
This gives a less dense species occurrence network, making hierarchical structures easier to detect.

\subsection*{Global mammal occurrence data}
\begin{figure}[htbp]
    \centering
    \includegraphics[width=0.6\textwidth]{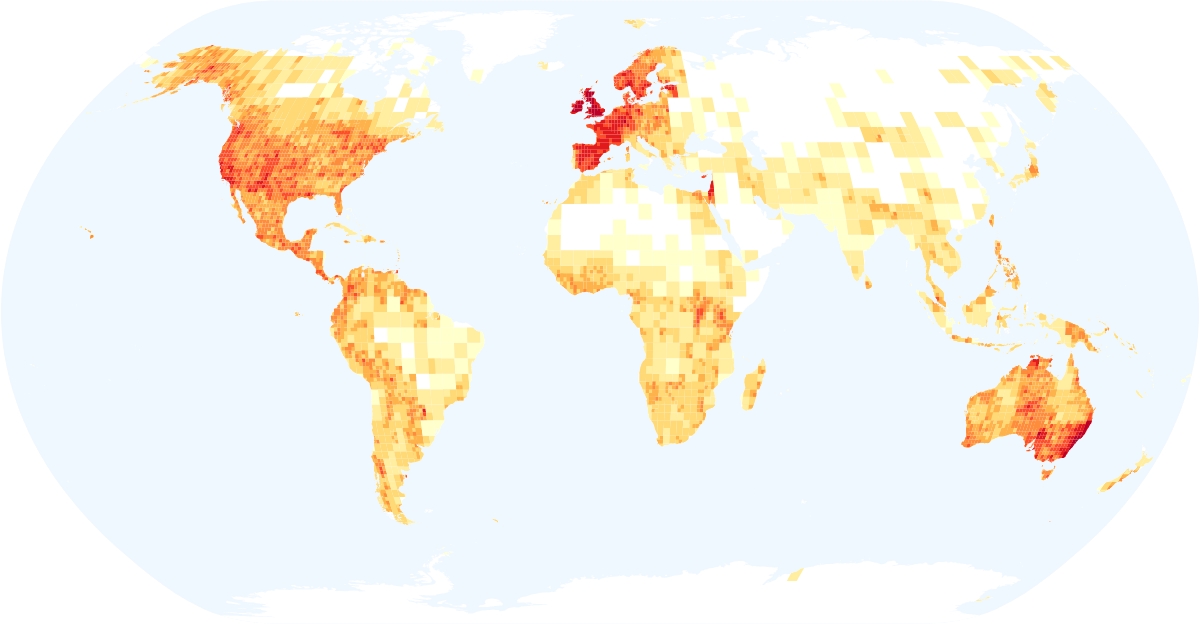}
    \caption{Mammal occurrences. It consists of 1.5M point occurrences of $4\,972$ species binned with an adaptive resolution to $5\,219$ grid cells with sides 1-4 degrees.}
    \label{fig:mammals-heatmap}
\end{figure}

To evaluate the method, we applied the same point occurrence dataset with globally distributed terrestrial mammals and the corresponding phylogenetic tree we used in~\cite{edler_infomap_2017}. It consists of 1.5M point occurrences that we bin with an adaptive resolution to grid cells from 4 degrees to 1, subdivided on capacity 100 with a minimum of 10 records per cell.
As a bipartite network without incorporating the phylogenetic tree, it has $10\,191$ nodes consisting of $4\,972$ species connected to $5\,219$ grid cells with $210\,892$ links weighted by range size. The species occurrences are not uniformly distributed but concentrated in Europe, sub-Saharan Africa, the Americas, and Australasia, with Russia being the most sparse region (Fig.~\ref{fig:mammals-heatmap}).

\section{Results}

\begin{figure}[htbp]
     \begin{subfigure}[t]{0.49\textwidth}
         \centering
         \includegraphics[width=\textwidth]{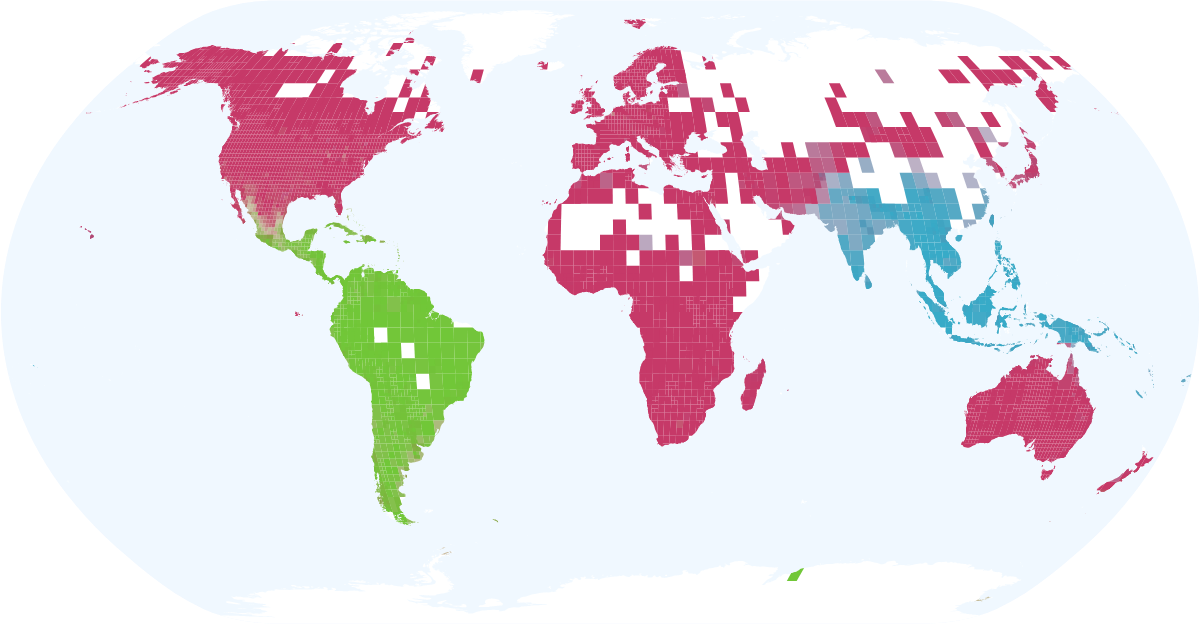}
         \caption{Level 1}
         \label{fig:mammals-1}
     \end{subfigure}
     \begin{subfigure}[t]{0.49\textwidth}
         \centering
         \includegraphics[width=\textwidth]{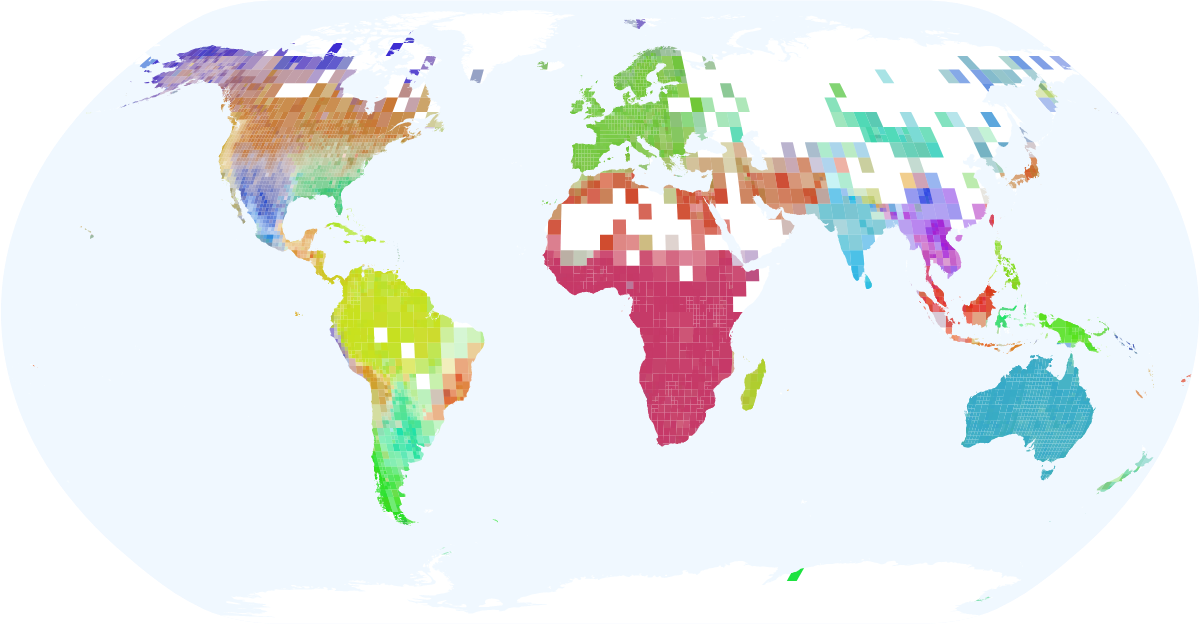}
         \caption{Level 2}
         \label{fig:mammals-2}
     \end{subfigure}
     \begin{subfigure}[t]{1\textwidth}
         \centering
         \includegraphics[width=0.7\textwidth]{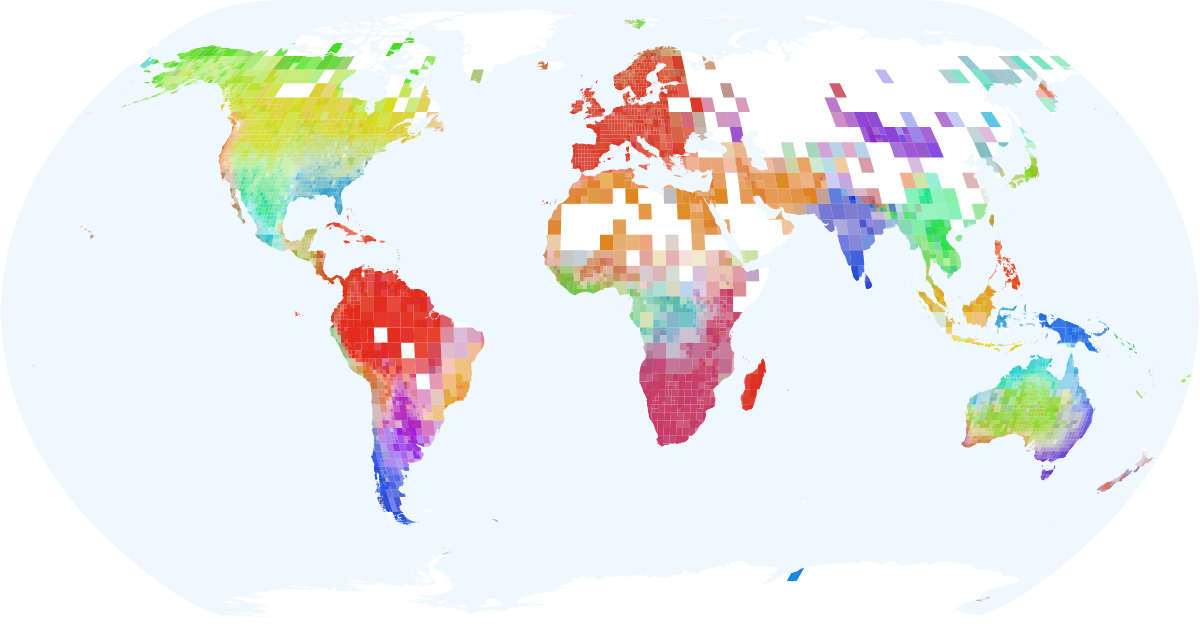}
         \caption{Level 3}
         \label{fig:mammals-3}
     \end{subfigure}
    \caption{Three levels of nested mammalian bioregions detected with Infomap Bioregions 2. Map based on 1.5M point occurrences binned with an adaptive resolution to grid cells from 4 degrees to 1. Opacity shows the degree of species overlap, which highlights transition zones}
    \label{fig:mammals}
\end{figure}

To test our method, we load the mammal dataset described above into Infomap Bioregions 2, with the default settings for adaptive resolution between 1 and 4 degrees.
As a baseline, we run Infomap with 20 trials without including the phylogenetic tree. The resulting bioregions are organized into three levels, with the top level containing tree modules. North America, Europe, Africa, North Asia, and Australia are grouped, while South America and South Asia are separated at this level (\Cref{fig:mammals-1}).
In the second level, at higher resolution, we see how Europe, Africa, North America, and Australia are divided. We also see how Asia and the North and South Americas are divided into more biogeographical units. Sub-Saharan Africa is kept in a single module, but Madagascar is distinctly separated (\Cref{fig:mammals-2}).
Australia and Africa are divided into finer bioregions at the finest level with clear transition zones (\Cref{fig:mammals-3}).

\begin{figure}[htbp]
    \begin{subfigure}[t]{0.24\textwidth}
        \centering
        \includegraphics[width=\textwidth]{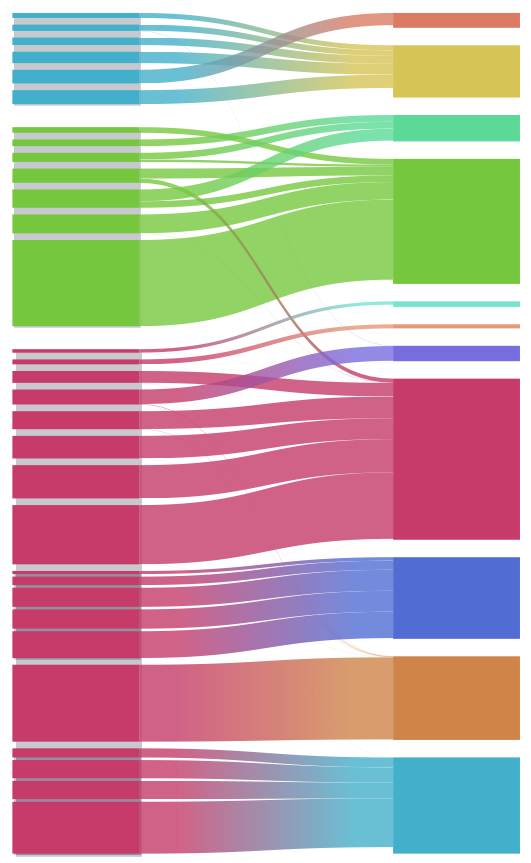}
         \caption{}
         \label{fig:mammals-tree-alluvial}
     \end{subfigure}
     \hfill
     \begin{subfigure}[t]{0.76\textwidth}
         \centering
        \includegraphics[width=\textwidth]{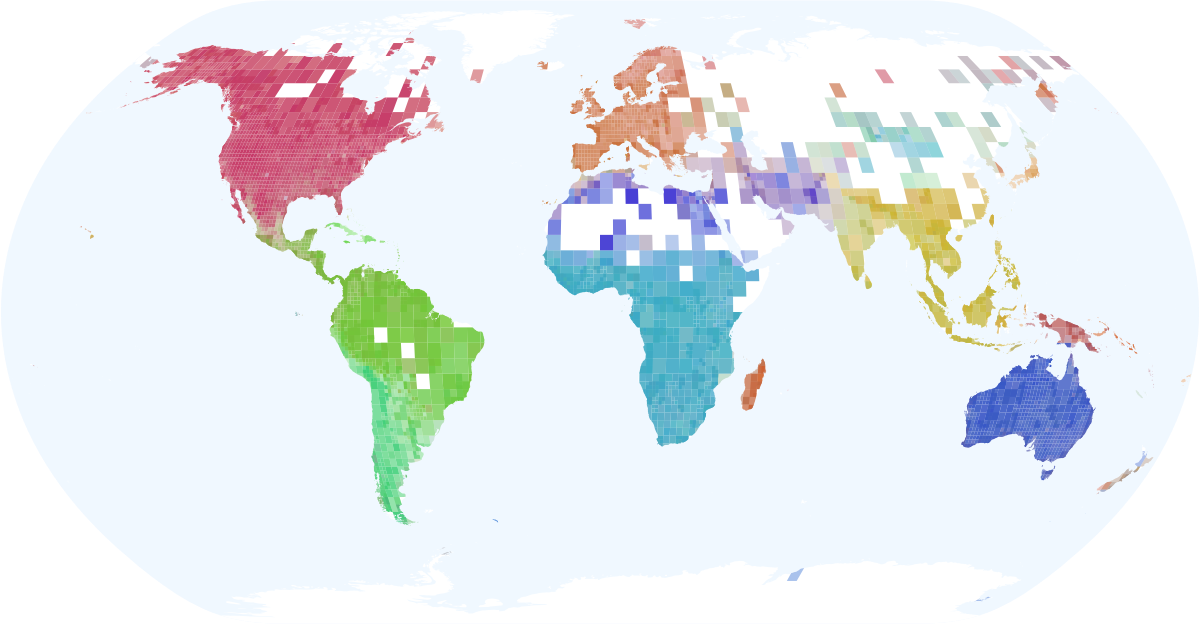}
         \caption{}
         \label{fig:mammals-tree-b}
     \end{subfigure}
    \caption{Comparing bioregions with and without integrating the phylogenetic tree. a) Alluvial diagram comparing the bioregions from \Cref{fig:mammals}, colored by the top level bioregions with larger and smaller gaps showing levels 2 and 3, respectively, with the bioregions obtained with the tree. b) One level of bioregions detected when integrating the whole phylogenetic tree, which tends to merge areas where species are phylogenetically close.}
    \label{fig:mammals-tree}
\end{figure}

When we integrate the whole phylogenetic tree of mammals, the network becomes denser, and we find bioregions in a single level. By connecting grid cells not only based on shared species but also shared ancestry from the whole tree, the bioregions capture broad phylogenetically more distinct areas  (\Cref{fig:mammals-tree}). We use an Alluvial diagram to compare the map without a tree (left) with the map based on the tree (right). In the left part, colors show bioregions on level 1, and larger and smaller vertical gaps show bioregions on levels 2 and 3, respectively. With the phylogenetic tree integrated, the map is largely contained within the top-level bioregions without a tree, where phylogenetically distinct bioregions from level 2, such as Madagascar and New Guinea, are kept separate while other bioregions are merged. Interestingly, bioregions produced when integrating phylogenetic information are more congruent with those proposed by Wallace and later refinements also using phylogenetic information~\cite{holt_update_2013}, supporting the validity of our approach.   

\begin{figure}[htbp]
    \centering
    \includegraphics[width=0.8\textwidth]{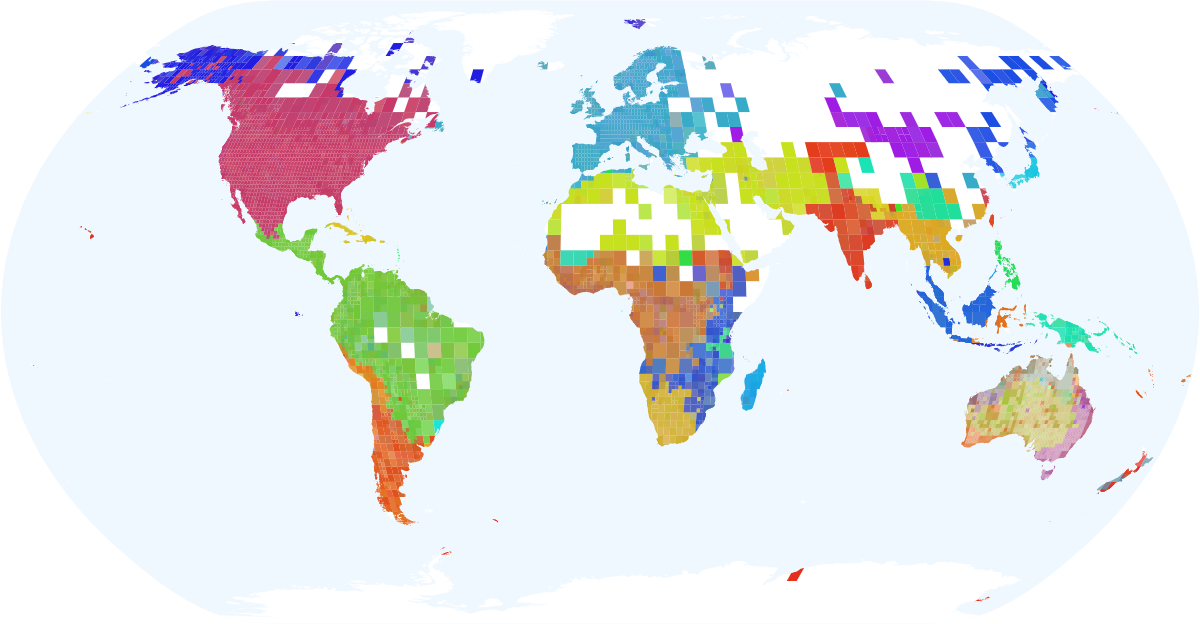}
    \caption{Overlapping bioregions. By segregating the network between species and grid cells at time 110Ma, using higher-order networks, we can detect evolutionarily distinct bioregions and see where they overlap, such as in Australia, based on the opacity and mixed colors. Showing the bottom level of four from a hierarchical result of nested bioregions.}
    \label{fig:mammals-seg}
\end{figure}

While integrating the whole phylogeny produce highly phylogenetically distinct bioregions, it may also blur the signal of to some extent recent event on the spatial organization of lineages. Around 110Ma the Earth experienced a global plate reorganization event with break-ups leading to the current configuration of continents~\cite{olierook_timing_2020}. Hence, it could be expected that important reconfigurations of biodiversity took place after this event, shaping the current distribution of lineages and leaving detectable signals on bioregions. To explore this idea, we segregate the network between species and grid cells at 110Ma, which creates bioregions formed only by the shared species within the evolutionary branches up to that point. This creates evolutionary distinct bioregions that may also overlap in space.  The overlap is highlighted with a lower opacity and mixed colors. The resultant bioregions are similar to those detected when using the whole phylogeny, though in the finest of four levels there are more detailed bioregions such as those dividing Africa (\Cref{fig:mammals-seg}). At this hierarchical level,  bioregions are relatively distinct except for Australia, which contains many overlapping bioregions and suggests that the phylogenetic signal behind this bioregion predates the age of segregation used (i.e. the ancestors of the Australian species diversified earlier than 110Ma.). Overall, these results show how this segregation approach can help to better understand the origin of bioregions. 

%When we integrate the whole phylogenetic tree, the network becomes denser, and we find bioregions in a single level. By connecting grid cells not only based on shared species but also shared ancestry from the whole tree, the bioregions capture broad phylogenetically more distinct areas  (\Cref{fig:mammals-tree}). We use an Alluvial diagram to compare the map without a tree (left) with the map based on the tree (right). In the left part, colors show bioregions on level 1, and larger and smaller vertical gaps show bioregions on levels 2 and 3, respectively. With the phylogenetic tree integrated, the map is largely contained within the top-level bioregions without a tree, where phylogenetically distinct bioregions from level 2, such as Madagascar and New Guinea, are kept separate while other bioregions are merged.

\clearpage

\section{Conclusion}
We have introduced a method to incorporate phylogenetic information into network-based bioregionalization to produce more biologically informative and robust bioregions. The method indirectly connects grid cells through not only shared species but also shared ancestry. It reduces biases caused by disagreements in species delimitation and also alleviates the risk of overfitting on sparse data where different but closely related species may end up in disconnected bioregions spanning a single or few cells.

We now weigh links between species and grid cells by the species’ geographic distribution which reduces the sensitivity on widespread species to overshadow spatial signals from other species and collapse modular patterns. Doing the same for ancestral nodes enables the integration of the whole phylogenetic tree. This highlights bioregions shaped by range-restricted species or clades, essential for conservation purposes and a better understanding of evolutionary processes shaping bioregional patterns. This generally also gives deeper hierarchies of nested bioregions which we now make possible to explore. Coloring the grid cells by the degree to which they connect to other bioregions highlights transition zones, fuzzy boundaries and the robustness of the solution on grid cell level.

For a more detailed analysis, we present two complementary methods to delineate evolutionary distinct bioregions by extracting information from a phylogenetic tree at selected times into the network. The first method has an integrating effect, connecting phylogenetic nodes at a selected time to grid cells where their descendant species occur. This procedure indirectly connects grid cells with species sharing common ancestors within the selected period. By sweeping through the phylogenetic tree in time, our method interpolates between finding bioregions based only on distributional data and finding spatially segregated clades, uncovering evolutionarily distinct bioregions at different time slices. The second method has a segregating effect, constraining network paths within branches cut at a selected point in time. With the segregating method, we can interpolate between finding evolutionary isolated clades and the spatial extent of clades, relaxing the non-overlapping constraint for bioregions.

We have implemented these methods in Infomap Bioregions version 2, an interactive web application that makes it easy to explore the relationship between species' spatial and phylogenetic patterns and identify evolutionarily distinct overlapping bioregions. With these new techniques, we can create richer bioregional maps and uncover biogeographical and evolutionary patterns not previously visible.
    
%Our method can incorporate other types of metadata besides evolutionary trees. A possible research direction is to include shared beneficial species or climatic variables.

%\section{Data availability}

%\section{Acknowledgements}

\section{Funding}

D.E.\ and M.R.\ were supported by the Swedish Research Council (2016-00796).
A.H.\ was supported by the Swedish Foundation for Strategic Research, Grant No. SB16-0089.
A.R.\ was partially supported by the Kone Foundation funded project Comparing evolutionary processes in nature and society (project number 202007064).
J.C.\ was supported by the Ministry of Science and Innovation through the UNIPER project (PID2020-114851GA-I00).
A.A.\ acknowledges financial support from the Swedish Research Council (2019-05191), the Swedish Foundation for Strategic Environmental Research MISTRA (Project BioPath), and the Royal Botanic Gardens, Kew.

\section{Author contributions}

A.A.\ conceived the study. D.E.\ designed the methods.
D.E.\ and A.H.\ implemented the interactive web application and performed the experiments with feedback from all authors.
All authors wrote, edited, and accepted the manuscript in its final form.

\section{Competing interests}

The authors declare that they have no competing interests.

\printbibliography

\clearpage
\appendix
\section*{Supplementary Information}

\setcounter{figure}{0}
\renewcommand{\thefigure}{S\arabic{figure}}

%\begin{figure}[htbp]
%    \centering
%    \includegraphics[width=0.8\textwidth]{figures/mammals_global_t_tw0.5_st0.7_l5.png}
%    \caption{Overlapping bioregions. Global mammals with tree and segregation time (6 levels).}
%    \label{fig:mammals-tree-seg}
%\end{figure}

\begin{figure}[htbp]
     \begin{subfigure}[t]{\textwidth}
         \centering
         \includegraphics[width=0.75\textwidth]{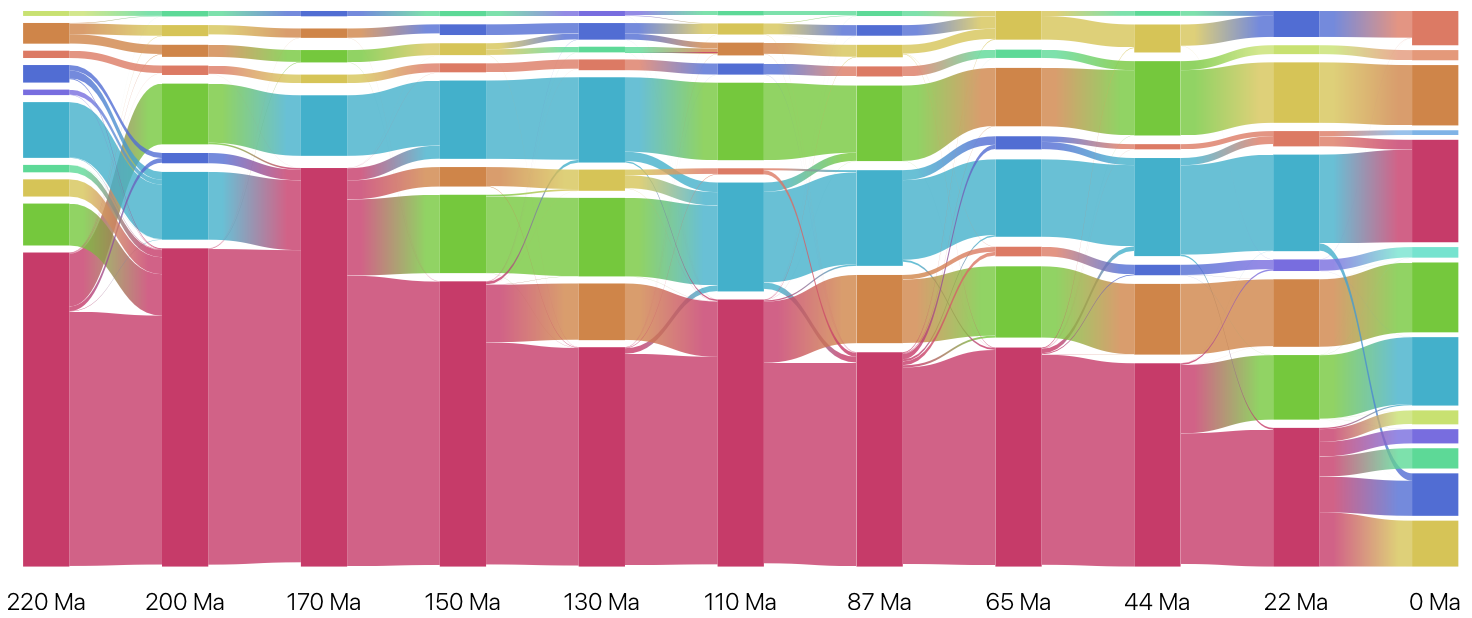}
         \caption{}
     \end{subfigure}
     \begin{subfigure}[t]{0.3\textwidth}
         \centering
         \includegraphics[width=\textwidth]{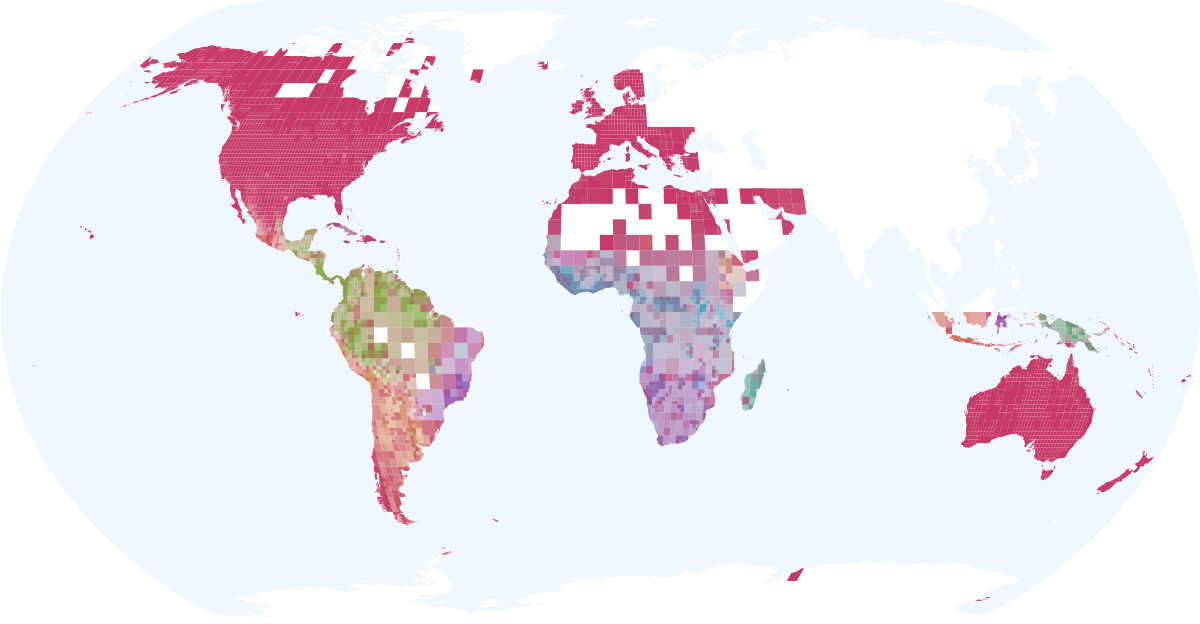}
         \caption{220 Ma}
     \end{subfigure}
     \hfill
     \begin{subfigure}[t]{0.3\textwidth}
         \centering
         \includegraphics[width=\textwidth]{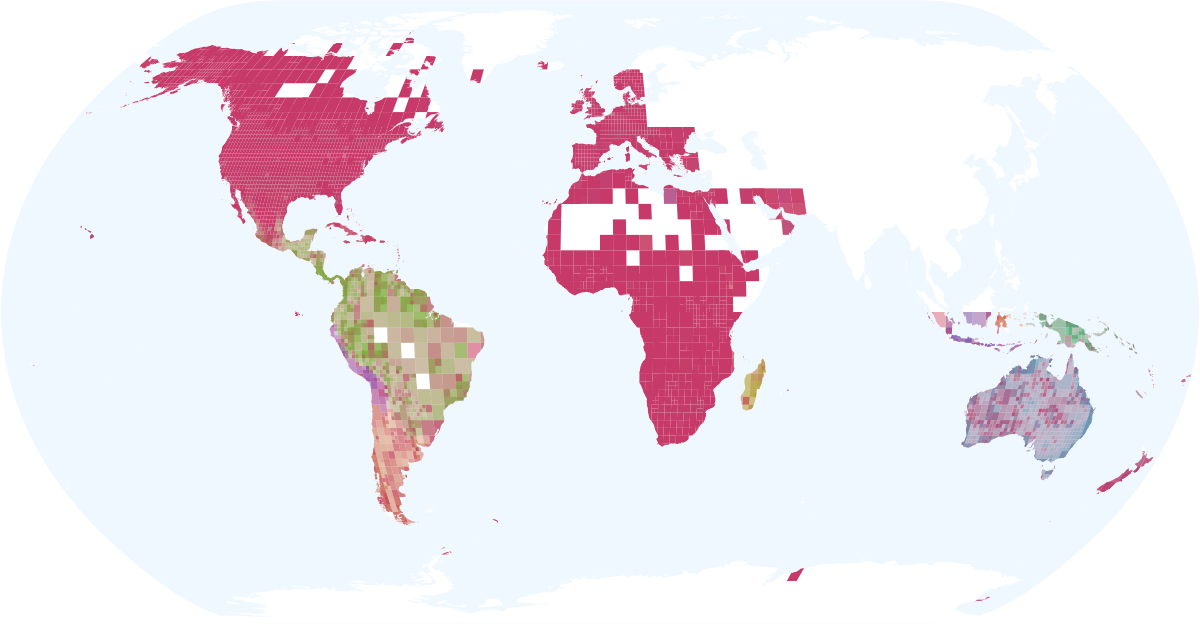}
         \caption{200 Ma}
     \end{subfigure}
     \hfill
     \begin{subfigure}[t]{0.3\textwidth}
         \centering
         \includegraphics[width=\textwidth]{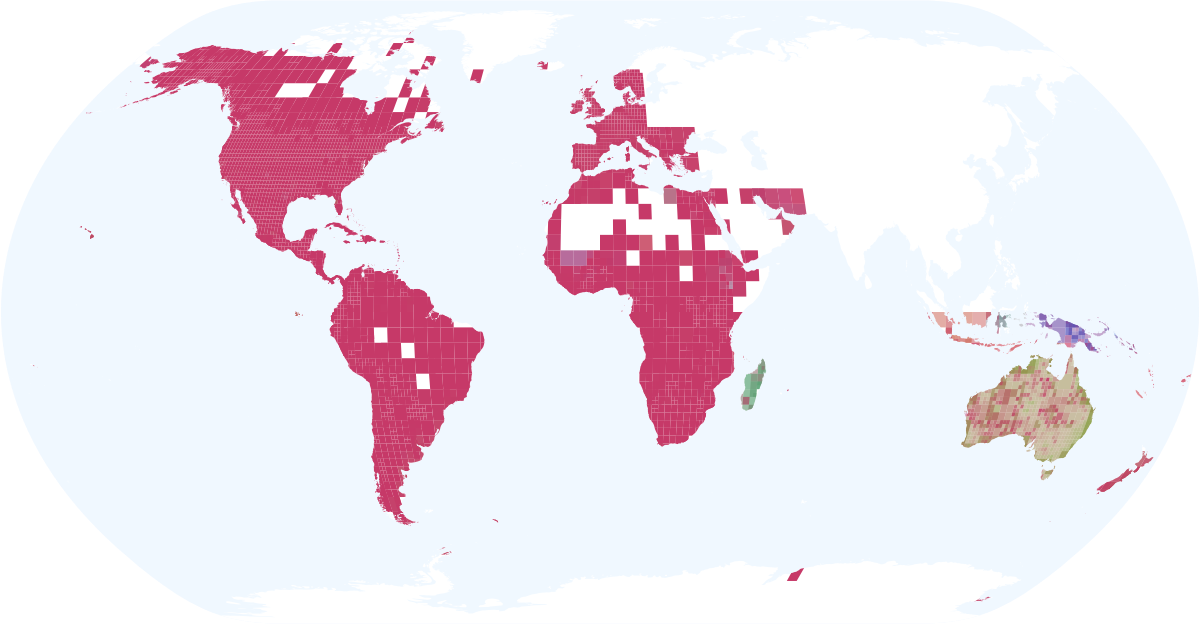}
         \caption{170 Ma}
     \end{subfigure}
     \hfill
     \begin{subfigure}[t]{0.3\textwidth}
         \centering
         \includegraphics[width=\textwidth]{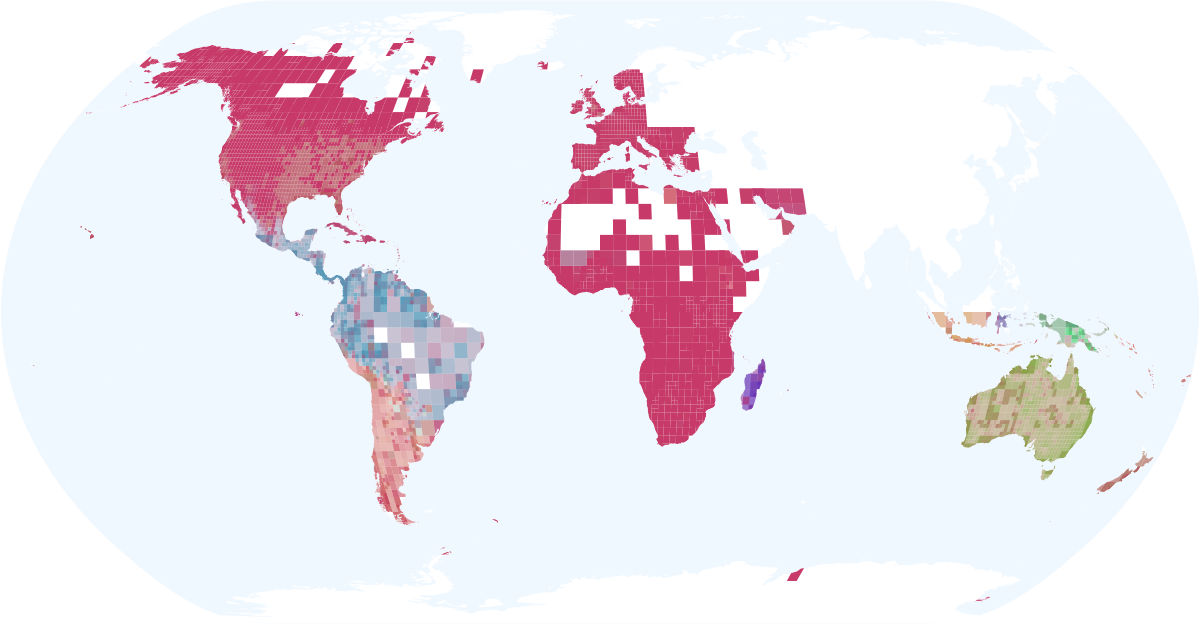}
         \caption{150 Ma}
     \end{subfigure}
     \hfill
     \begin{subfigure}[t]{0.3\textwidth}
         \centering
         \includegraphics[width=\textwidth]{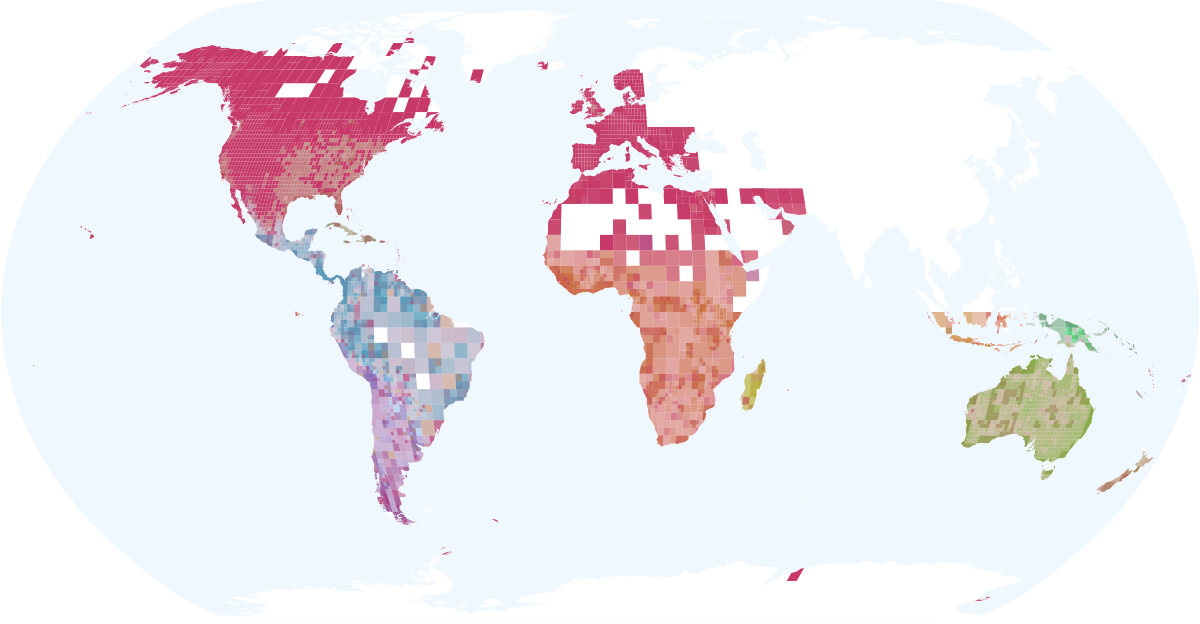}
         \caption{130 Ma}
     \end{subfigure}
     \hfill
     \begin{subfigure}[t]{0.3\textwidth}
         \centering
         \includegraphics[width=\textwidth]{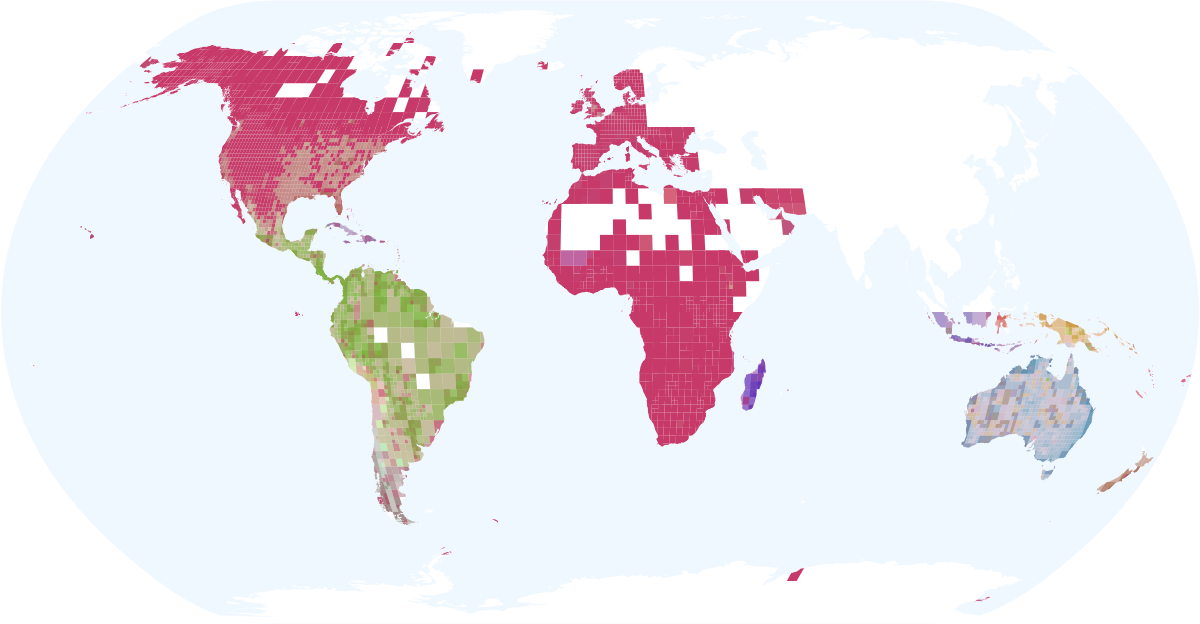}
         \caption{110 Ma}
     \end{subfigure}
     \hfill
     \begin{subfigure}[t]{0.3\textwidth}
         \centering
         \includegraphics[width=\textwidth]{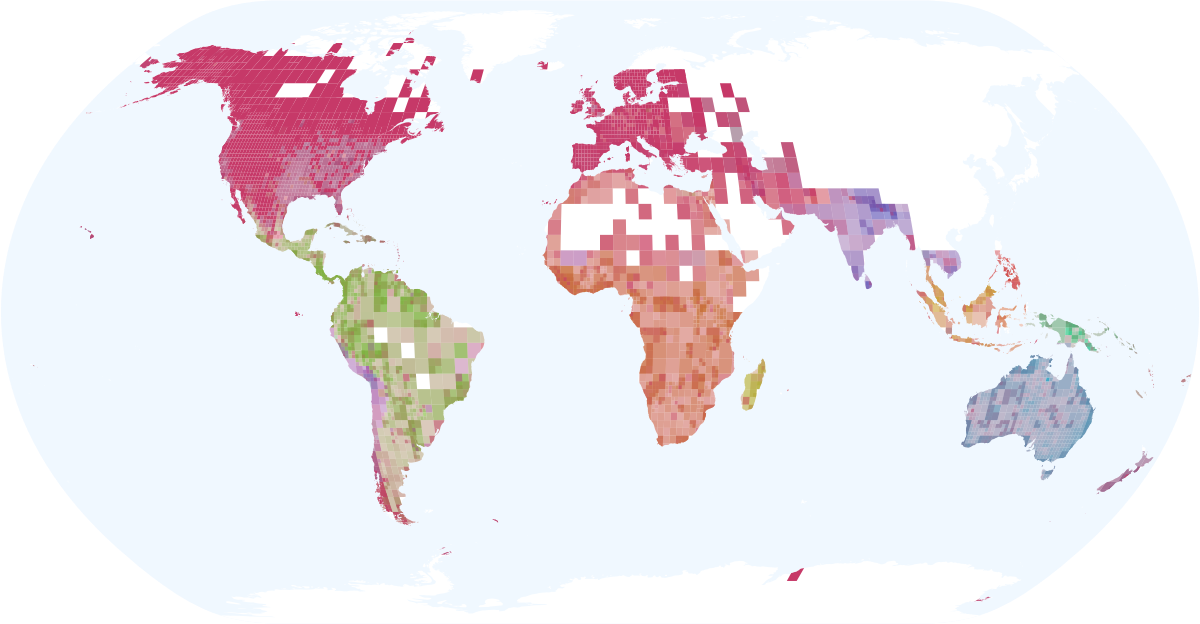}
         \caption{87 Ma}
     \end{subfigure}
     \hfill
     \begin{subfigure}[t]{0.3\textwidth}
         \centering
         \includegraphics[width=\textwidth]{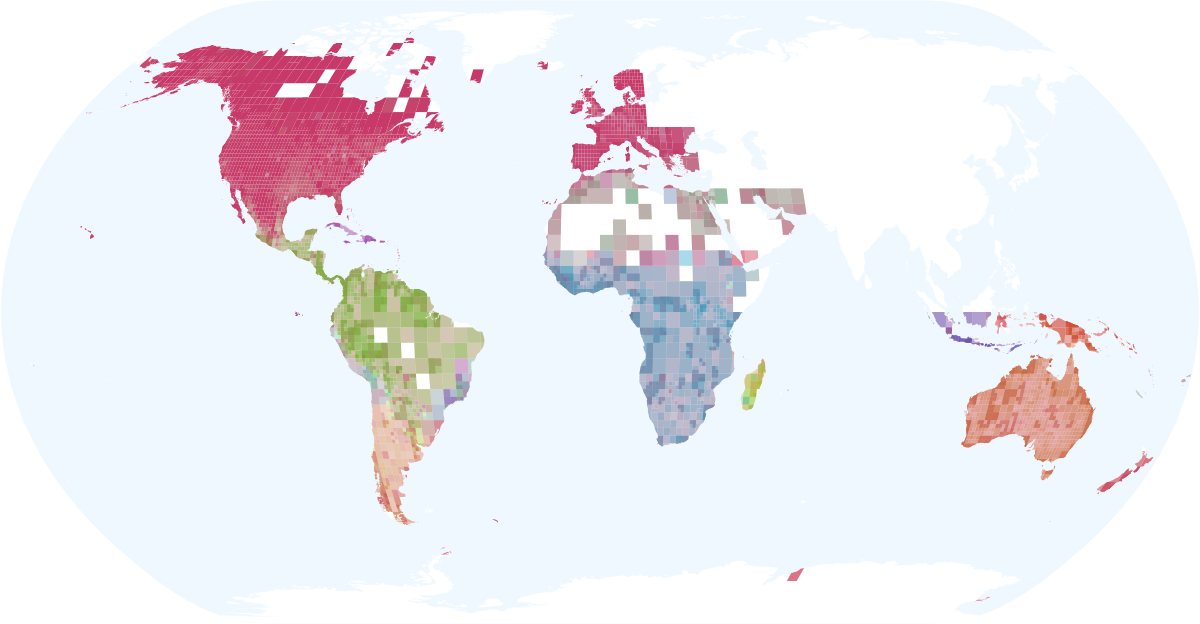}
         \caption{65 Ma}
     \end{subfigure}
     \hfill
     \begin{subfigure}[t]{0.3\textwidth}
         \centering
         \includegraphics[width=\textwidth]{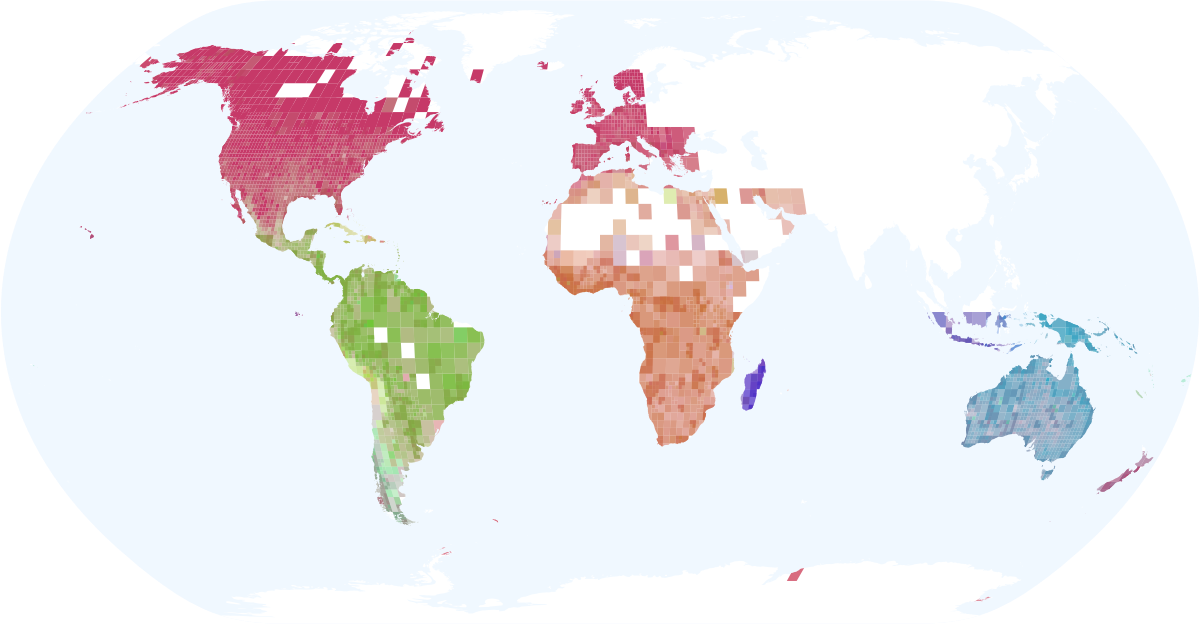}
         \caption{44 Ma}
     \end{subfigure}
     \hfill
     \begin{subfigure}[t]{0.3\textwidth}
         \centering
         \includegraphics[width=\textwidth]{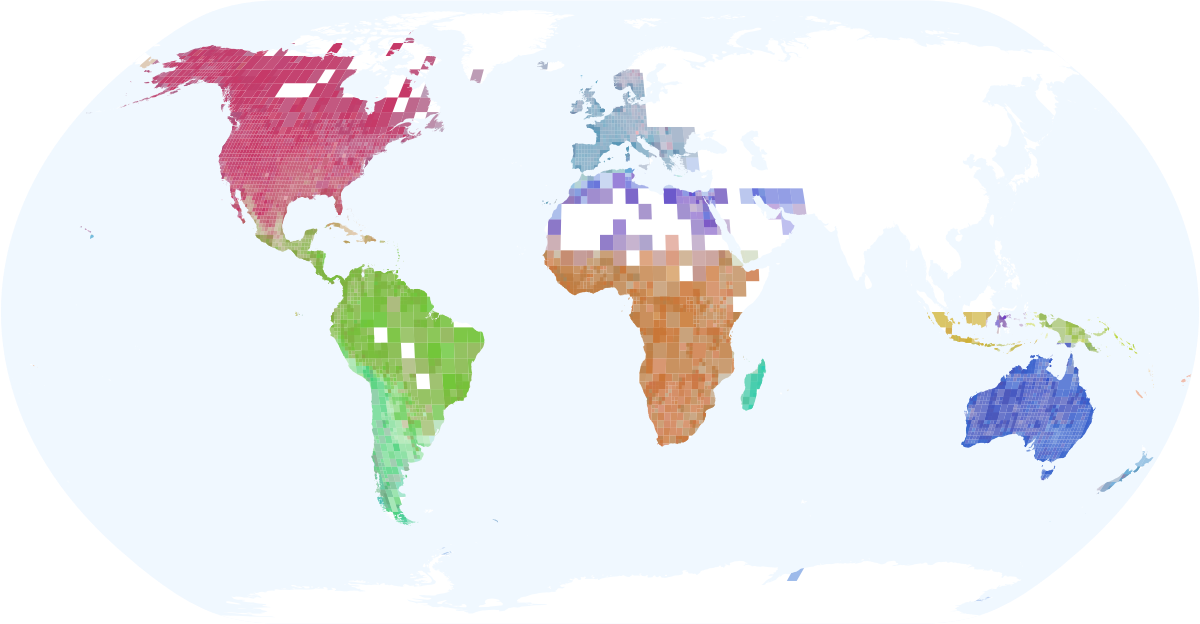}
         \caption{22 Ma}
     \end{subfigure}
     \hspace{1.5em}
     %\hfill
     \begin{subfigure}[t]{0.3\textwidth}
         \centering
         \includegraphics[width=\textwidth]{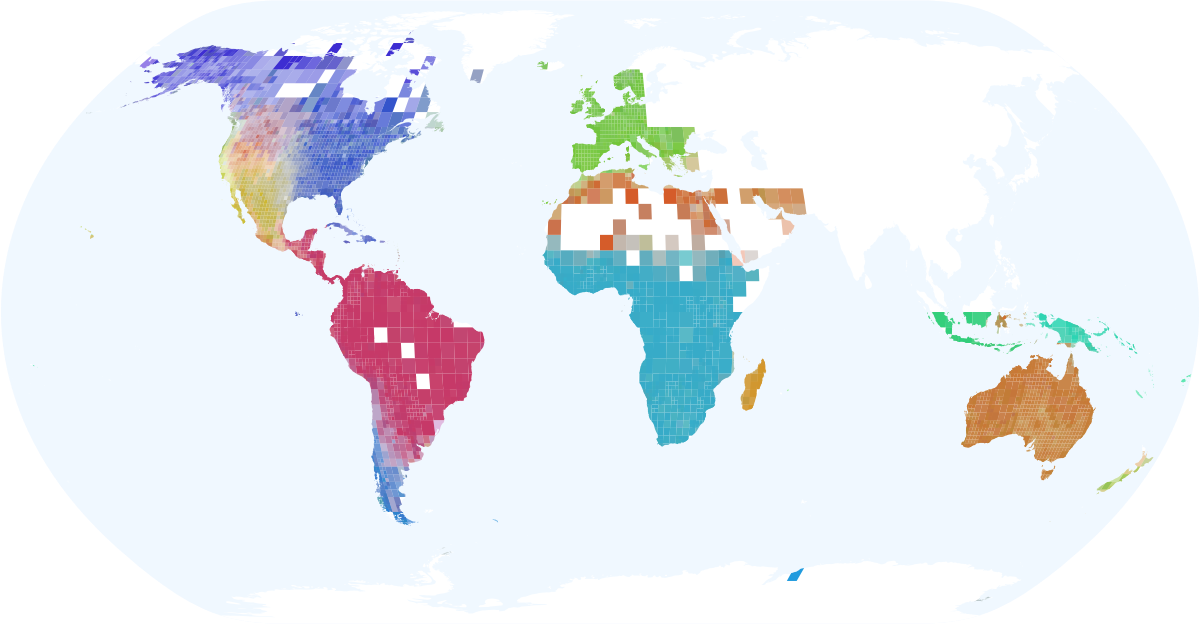}
         \caption{0 Ma}
     \end{subfigure}
    \caption{Evolutionarily unique mammalian bioregions. We integrate the ancestral nodes at a selected time with equal strength as the species and sweep through the time from recent to 220Ma. As we go back in time, grid cells are connected by shared species at that time. If the species in two bioregions to a large degree share a common ancestor within the selected time, two bioregions are likely merged at that point. More and more regions get merged, leaving evolutionarily unique areas left such as Madagascar. }
    \label{fig:mammals-sweep}
\end{figure}

\end{document}